\documentclass[fleqn,usenatbib]{mnras}

\usepackage{newtxtext,newtxmath}

\usepackage[T1]{fontenc}

\DeclareRobustCommand{\VAN}[3]{#2}
\let\VANthebibliography\thebibliography
\def\thebibliography{\DeclareRobustCommand{\VAN}[3]{##3}\VANthebibliography}

\usepackage{graphicx}	
\usepackage{amsmath}	

\title[The Neptunian ridge is a matter of scale]{The Neptunian ridge depends on the choice of linear or logarithmic period}

\author[K. Cui et al.]{
Kaiming Cui,$^{1,2}$\thanks{E-mail: Kaiming.Cui@warwick.ac.uk}
and David J. Armstrong$^{1,2}$
\\
$^{1}$Department of Physics, University of Warwick, Gibbet Hill Road, Coventry CV4 7AL, UK\\
$^{2}$Centre for Exoplanets and Habitability, University of Warwick, Gibbet Hill Road, Coventry CV4 7AL, UK\\
}

\date{Accepted XXX. Received YYY; in original form ZZZ}

\pubyear{\the\year{}}

\begin{document}
\label{firstpage}
\pagerange{\pageref{firstpage}--\pageref{lastpage}}
\maketitle

\begin{abstract}
The Neptunian ridge has been reported as an overdensity of planets with radii of $\simeq$5--8\,R$_\oplus$ and orbital periods of $\simeq$3--6\,d, located between the Neptunian desert and the savanna. It was first identified in Kepler data \citep{castro-gonzalezMappingExoNeptunianLandscape2024} and has recently been reanalysed using our TESS occurrence rates \citep{castro-gonzalezNeptunianRidgeLens2026a}, where it appears as an approximately twofold excess relative to neighbouring regions.  
We show that the reported excess arises from expressing the occurrence density per unit period, $\mathrm{d}\,N/\mathrm{d}P$, rather than per unit logarithmic period, $\mathrm{d}\,N/\mathrm{d}\ln P$, where the logarithmic form is standard in planet demographics. Because the savanna interval used for comparison is twice as wide in period as the ridge, a population uniform in $\ln P$ naturally produces a factor-of-two contrast in $\mathrm{d}\,N/\mathrm{d}P$. When expressed per unit $\ln P$, the published TESS and Kepler ridge-to-savanna contrasts decrease from 1.6--2.2 to 0.8--1.2, all consistent with unity. A simulated population with a desert but no intrinsic excess reproduces the appearance of a ridge in $\mathrm{d}\,N/\mathrm{d}P$ precisely.  
Reanalysing the TESS-SPOC occurrence rates of 5.5--8.5\,R$_\oplus$ planets around FGK stars, we find a ridge-to-savanna ratio of $0.91^{+0.21}_{-0.16}$ per unit $\ln P$, which does not support the claimed overdensity at the ridge. Taken together, these tests imply that the published ridge is a result of the scale choice used, rather than a definitive property of the observed planet population.
\end{abstract}




\section{Introduction}
The Neptunian ridge is an overdensity of planets with $5.5<\mathrm{R_p}<8.5\,\mathrm{R_\oplus}$ at $3.2\la P\la5.7$\,d, first identified by \citet[][hereafter CG24]{castro-gonzalezMappingExoNeptunianLandscape2024} using Kepler DR25 planets weighted by their transit detection probabilities. It is located between the Neptunian desert \citep[e.g.,][]{szaboShortperiodCensorSubJupiter2011,mazehDearthShortperiodNeptunian2016} and the savanna \citep{bourrierDREAMOrbitalArchitecture2023}.

In the second paper of our TESS demographics series \citep[][hereafter Paper II]{cuiDemographicsCloseInTESS2026a}, we systematically measured close-in occurrence rates around F-, G-, K-type main-sequence stars from four years of TESS-SPOC full-frame image data following the pipeline of \citet[][hereafter Paper I]{cuiDemographicsCloseinTESS2026}. One of our results showed that the occurrence profile of 6--8$\mathrm{R_\oplus}$ planets can be explained by a smooth broken power rather than requiring a significant log-normal component to depict the overdensity, leading to the conclusion that a distinct Neptunian ridge is not required by the TESS data. 

Analysing the occurrence posteriors released with Paper I and II, \citet[][hereafter CG26]{castro-gonzalezNeptunianRidgeLens2026} reach the opposite conclusion: for planets of $\simeq$5--8\,$R_\oplus$, the ridge has twice the occurrence density of the denser neighbouring region, with 99.9 per cent posterior support (a one-sided Gaussian equivalent of $3.1\sigma$).

In this work, we show that the two conclusions differ mainly because of the variable in which the occurrence rate density is expressed. CG24 and CG26 analyse the density per unit period $\mathrm{d}\,N/\mathrm{d}P$, whereas we, like almost all other occurrence-rate studies, use the density per unit logarithmic period, $\mathrm{d}\,N/\mathrm{d}\ln P$.

\section{Linear vs. logarithmic period}

\begin{figure*}
    \centering
    \includegraphics[width=\linewidth]{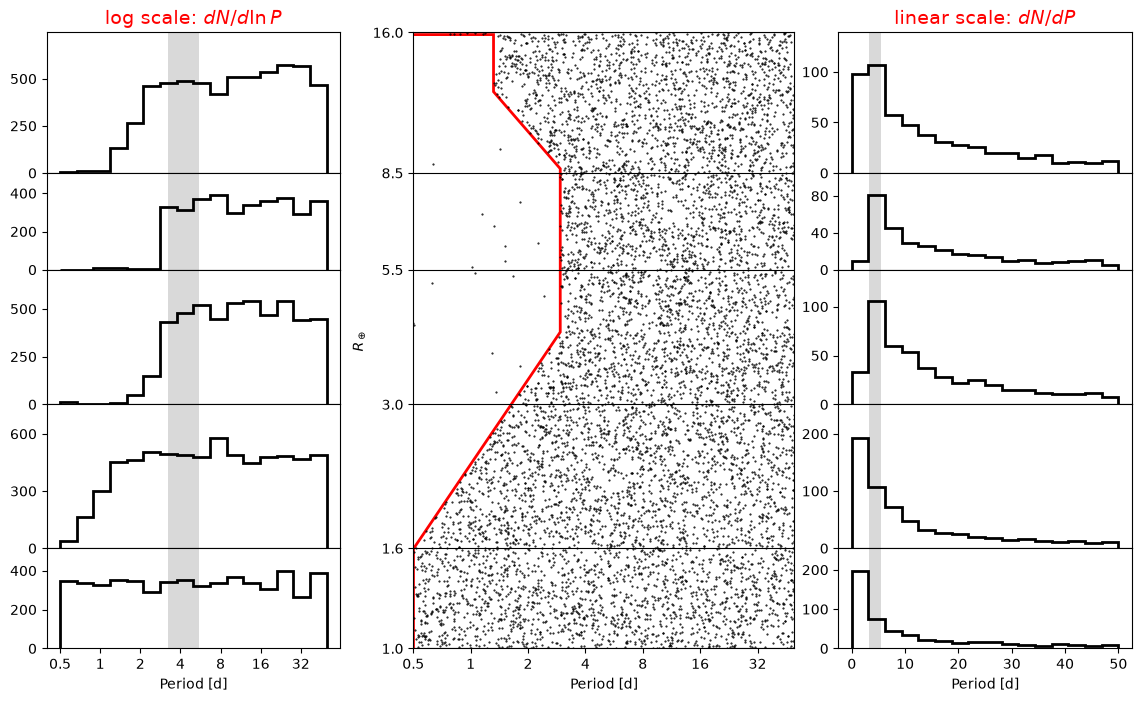}
    \caption{A simulated population with a desert but no ridge. Center: 10,000 planets drawn uniformly in $\ln P$ and $\ln R_\mathrm{p}$, with the population depleted by a factor of 100 inside the Neptunian desert boundaries of CG24 (red lines). Horizontal lines delineate five radius strips. Left: the period distribution in each strip, shown in equal bins of $\ln P$ and proportional to $\Gamma=\mathrm{d}N/\mathrm{d}\ln P$. Right: the same strips, shown in equal bins of $P$ and proportional to $h=\mathrm{d}N/\mathrm{d}P$. The gray bands mark the ridge period range identified by CG24, 3.2--5.7\,d. Although $\Gamma$ is flat outside the desert, the strips whose desert edge lies at periods of a few days exhibit a ridge-like peak in $h$; the 5.5--8.5\,$R_\oplus$ strip corresponds to the ridge reported by CG24.}
    \label{fig:sim}
\end{figure*}

\subsection{Definition of occurrence rate density}
\label{sec:def} 
Planet occurrence rate density is usually described by the number of planets per star per unit logarithm of the orbital period,
\begin{equation}\label{eq:log}
    \Gamma(P) \equiv \frac{\mathrm{d}\,N}{\mathrm{d}\ln P},
\end{equation}
which is the quantity used to study many important characteristics, for example, hot and cold giant planets, sub-Neptunes, the radius valley, and the Neptunian desert \citep[e.g.,][]{cummingKeckPlanetSearch2008,howardPLANETOCCURRENCE0252012a,fressinFALSEPOSITIVERATE2013,mazehDearthShortperiodNeptunian2016,fultonCaliforniaKeplerSurveyIII2017,petiguraCaliforniaKeplerSurveyIV2018}, and the quantity we used in Papers I and II. However, CG24 and CG26 instead describe the period distribution by the density per unit period,
\begin{equation}\label{eq:lin}
    h(P) \equiv \frac{\mathrm{d}\,N}{\mathrm{d}\,P}.
\end{equation}
The two are related by a factor of $P$.
\begin{equation}\label{eq:jacobian}
    \Gamma(P) = P\,h(P),
\end{equation}
thus a population that is uniform in $\ln P$ ($\Gamma$ = constant) has $h\propto P^{-1}$, and a power law $\Gamma\propto P^{\alpha}$ becomes $h\propto P^{\alpha-1}$. Some features of a distribution can survive with this transformation, e.g., a sharp edge remains an edge, and a profile that departs from a single power law in one variable also does so in the other. However, a local maximum does not survive in general. Where $\Gamma$ rises steeply and then flattens, as it does at the edge of a desert, $h$ rises and then falls with $P^{-1}$, so that a peak appears in $h$ at the edge although $\Gamma$ has none.

For a ridge interval $R$ and a savanna interval $S$ with integrated occurrence $F_R$ and $F_S$, the contrasts between their mean densities in the two variables are
\begin{equation}\label{eq:contrasts}
    C \equiv \frac{F_R/\Delta P_R}{F_S/\Delta P_S}, \qquad
    C_{\ln} \equiv \frac{F_R/\Delta \ln P_R}{F_S/\Delta \ln P_S},
\end{equation}
where $\Delta P$ and $\Delta\ln P$ are the widths of each interval in period and in logarithmic period, and $C$ is the exterior contrast of CG26. A log-uniform population with no ridge has $C_{\ln}=1$ but $C=C_\mathrm{flat}$, where
\begin{equation}\label{eq:cflat}
    C_\mathrm{flat} = \frac{\Delta\ln P_R/\Delta P_R}{\Delta\ln P_S/\Delta P_S},
\end{equation}
and in general $C_{\ln}=C/C_\mathrm{flat}$. When both intervals span the same factor in period, $C_\mathrm{flat}=\Delta P_S/\Delta P_R$, which equals 2 for the ridge ($R=[2.83, 5.66]$\,d) and savanna ($S=[5.66, 11.31]$\,d) adopted by CG26. A log-uniform population therefore has twice the density per unit period in the ridge as in the savanna, matching the excess reported by CG26, $C=2.14^{+0.64}_{-0.49}$.

CG26 also compiled ridge--savanna contrasts from the Kepler occurrence studies of \citet{howardPLANETOCCURRENCE0252012a}, \citet{fressinFALSEPOSITIVERATE2013}, \citet{muldersStelLARMASSDEPENDENTDROPPLANET2015}, \citet{kunimotoSearchingEntiretyKepler2020}, and CG24 (their table D.2). We reproduce them in Table~\ref{tab:contrasts}, together with $C_\mathrm{flat}$ and $C_{\ln}$ for each pair of intervals. Every measured contrast is close to its log-uniform expectation: $C$ ranges from 1.58 to 2.16, while $C_\mathrm{flat}$ ranges from 1.71 to 2.06. Per unit logarithmic period, the contrasts fall to $C_{\ln}=0.80$--1.21, all consistent with unity within their uncertainties.
This also explains why no ridge was noted in the earlier Kepler studies, which reported occurrence rates in logarithmically spaced period bins, whereas CG24 identified it from weighted cumulative distributions and histograms in linear period.

\begin{table*}
    \caption{Ridge--savanna occurrence contrasts per unit period ($C$) and per unit logarithmic period ($C_{\ln}$), together with the contrast expected for a log-uniform population with no ridge ($C_\mathrm{flat}$, equation~(\ref{eq:cflat})).}
    \label{tab:contrasts}
    \renewcommand{\arraystretch}{1.25}
    \begin{tabular}{ccccccl}
      \hline
      Radius range ($R_\oplus$) & Ridge, $R$ (d) & Savanna, $S$ (d) & $C$ & $C_\mathrm{flat}$ & $C_{\ln}=C/C_\mathrm{flat}$ & Reference \\
      \hline
      5.66--8.00 & 3.42--5.85 & 5.85--10.00 & $1.58^{+1.11}_{-0.74}$ & 1.71 & $0.92^{+0.65}_{-0.43}$ & \citet{howardPLANETOCCURRENCE0252012a} \\
      4.00--6.00 & 3.42--5.85 & 5.85--10.00 & $2.07^{+1.20}_{-0.71}$ & 1.71 & $1.21^{+0.70}_{-0.42}$ & \citet{fressinFALSEPOSITIVERATE2013} \\
      5.50--8.50 & 3.20--5.70 & 5.70--13.20 & $1.65^{+0.64}_{-0.47}$ & 2.06 & $0.80^{+0.31}_{-0.23}$ & \citet{muldersStelLARMASSDEPENDENTDROPPLANET2015} \\ 
      4.00--8.00 & 3.13--6.25 & 6.25--12.50 & $2.08^{+1.62}_{-0.88}$ & 2.00 & $1.04^{+0.81}_{-0.44}$ & \citet{kunimotoSearchingEntiretyKepler2020} \\
      5.50--8.50 & 3.20--5.70 & 5.70--13.20 & $2.16^{+0.99}_{-0.73}$ & 2.06 & $1.05^{+0.48}_{-0.35}$ & CG24 \\
      5.02--7.96 & 2.83--5.66 & 5.66--11.31 & $2.14^{+0.64}_{-0.49}$ & 2.00 & $1.07^{+0.32}_{-0.25}$ & Paper~I \\
      \hline
      5.50--8.50 & 2.83--5.66 & 5.66--11.31 & $1.82^{+0.42}_{-0.32}$ & 2.00 & $0.91^{+0.21}_{-0.16}$ & This work (Section~\ref{sec:tess}) \\
      \hline
    \end{tabular}
    \par\smallskip
    \parbox{\linewidth}{\footnotesize \textit{Notes.} The first six rows are adapted from table D.2 of CG26; the first five rows are based on Kepler and the last two on TESS-SPOC. $C$ and $C_{\ln}$ are defined in equation~(\ref{eq:contrasts}).}
\end{table*}

\subsection{Simulation}\label{sec:sim}
To intuitively show the effect, we simulate a population that has a desert but no ridge. We draw 10,000 planets uniformly in $\ln P$ over 0.5--50\,d and in $\ln R_\mathrm{p}$ over 1--16\,$R_\oplus$, and keep only 1 per cent of those that fall inside the desert boundaries of CG24 (their equations 6--9), so that the desert is a hundred times less populated than its surroundings (based on our Neptunian desert occurrence rates results in Papers I and II). We then compute the period distribution in five radius strips, both per unit $\ln P$ and per unit $P$  (Fig.~\ref{fig:sim}). By construction, $\Gamma$ is flat outside the desert in every strip, with only a step at the desert edge. In $h$, however, every strip whose desert edge lies at a few days shows a peak just outside the edge followed by a $P^{-1}$ decline. 

In particular, the 5.5--8.5\,$R_\oplus$ strip, which spans the radius range of the ridge, reproduces the desert--ridge--savanna distribution of the Kepler histogram of CG24 (their fig.~2) and of the TESS profiles of CG26 (their fig.~1), although the simulated population contains no excess at the ridge. 
Quantitatively, applying the intervals of CG24 in Table D.2 of CG26 gives $C=2.14$ in this simulation, almost identical with their reported contrast $2.16^{+0.99}_{-0.73}$, indicating that the ridge can be explained solely by the desert and a log-uniform distribution expressed on a linear scale.


\subsection{Why choose a logarithmic scale?}
The choice of logarithmic scale is more than a convention. The processes thought to sculpt the planetary population depend on orbital period through power laws: Kepler's third law gives $a\propto P^{2/3}$, the bolometric insolation scales as $S\propto P^{-4/3}$; the tidal circularisation time scale scales as $\tau_\mathrm{circ}\propto a^{13/2}\propto P^{13/3}$ \citep{goldreichSolarSystem1966}; and the minimum orbital period for tidal survival depends on the mean planetary density as $P_\mathrm{Roche}\propto\rho_\mathrm{p}^{-1/2}$, etc.

Without a preferred scale, the natural reference distribution assigns equal numbers of planets to equal factors in period, that is, it is uniform in $\ln P$ over a limited range, and any excess, whether produced by high-eccentricity migration or otherwise, should be measured relative to it. 
A local maximum of $h$ instead tests against $h$ constant, that is $\Gamma\propto P$, a reference under which the savanna of CG26 would hold twice as many planets per unit $\ln P$ as the ridge. Because both the location and even the existence of a maximum depend on the variable used, any claimed overdensity should specify the variable in which it is measured and be tested against a smooth reference in that variable. 

Working in linear rather than logarithmic period would also reshape other well-established features. Small planets (2--4\,$R_\oplus$) become more common per unit $\ln P$ up to $P_0\simeq7$\,d and then flatten \citep{howardPLANETOCCURRENCE0252012a}; in $h$, this knee becomes a peak at $\simeq$9\,d, a sub-Neptune `ridge' that has never been interpreted as a distinct population. Giant-planet occurrence per unit $\ln P$ increases with period up to a turnover near the snow line \citep{cummingKeckPlanetSearch2008,fernandesHintsTurnoverSnow2019,fultonCaliforniaLegacySurvey2021}, whereas $h$ would decline monotonically from the hot-Jupiter pile-up without any turnover. The lower desert boundary of Paper~II, $R_\mathrm{p}\propto P^{0.50}$ as a contour of $\Gamma$, would become $R_\mathrm{p}\propto P^{0.24}$ as a contour of $h$. Moreover, the radius valley is a straight line in the $\ln P$--$\ln R_\mathrm{p}$ plane, $R_\mathrm{p}\propto P^{m}$ with $m\simeq-0.1$ \citep{vaneylenAsteroseismicViewRadius2018}, in linear $P$ it becomes a curve whose slope, $\mathrm{d}R_\mathrm{p}/\mathrm{d}P\propto P^{m-1}$, changes with period, so no single index describes it.

For these reasons, $\mathrm{d}\,N/\mathrm{d}\,\ln P$ is the standard choice in occurrence-rate studies. Only a few studies \citep{gaudiPeriodDistributionClosein2005a,burkeTERRESTRIALPLANETOCCURRENCE2015a,neilJointMassRadius2020} have described the period distribution per unit period rather than per unit $\ln P$, and these either retained the logarithmic variable in their main analysis or adopted a power-law baseline, which remains a power law under the change of variable, so the choice does not affect their conclusions. By contrast, CG24 and CG26 identify the ridge as a local maximum of $h$, which need not correspond to a maximum of $\Gamma$ (Section~\ref{sec:def}).

\section{TESS-SPOC occurrence of 5.5--8.5\,{$R_\oplus$} planets}\label{sec:tess}
\begin{figure}
    \centering
    \includegraphics[width=\linewidth]{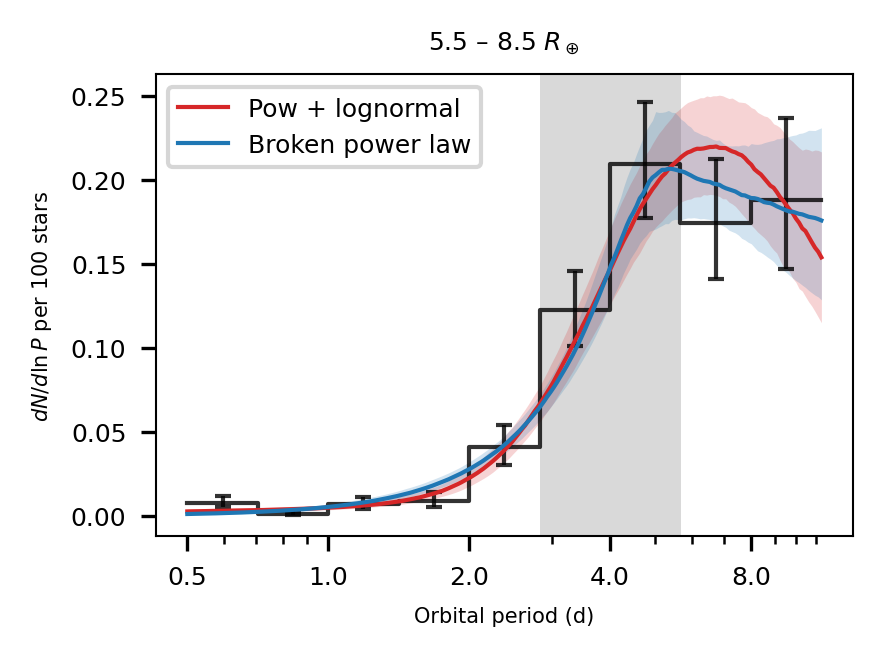}
    \caption{Occurrence-rate density per unit $\ln P$ of TESS-SPOC planets with $5.5<R_\mathrm{p}<8.5\,R_\oplus$ around FGK main-sequence stars. The histogram shows the binned estimates (posterior medians with 16th--84th percentile error bars). The red and blue curves show the power-law-plus-log-normal and broken-power-law models, with shaded $1\sigma$ credible bands. The grey band marks the ridge interval adopted by CG26 and in Table~\ref{tab:contrasts}, 2.83--5.66\,d.}
    \label{fig:tess}
\end{figure}

To compare directly with CG24 and CG26, we re-measured the one-dimensional occurrence profile with the TESS-SPOC FGK sample, using detection pipeline and hierarchical Bayesian framework of Papers~I and II, restricted to the radius range in which CG24 identified the ridge, $5.5<R_\mathrm{p}<8.5\,R_\oplus$, and to $0.5<P<11.31$\,d, which covers the desert, ridge, and savanna intervals like CG26 did. Fig.~\ref{fig:tess} shows $\Gamma$ in bins of a factor of $\sqrt{2}$ in period. The occurrence rises by more than an order of magnitude from 1.41 to 5.66\,d and is flat, within the uncertainties, from 4 to 11.31\,d. The ridge interval contains the upper part of this rise: its inner bin (2.83--4.00\,d) lies below both savanna bins.

Using two bins for each region, as CG26 did, we find a ridge-to-savanna ratio of $C_{\ln}=0.91^{+0.21}_{-0.16}$, with only 33 per cent of the posterior samples supporting an excess (Table~\ref{tab:contrasts}). Even if the ridge is represented by its most populated bin alone (4.00--5.66\,d), the ratio is $1.15^{+0.30}_{-0.23}$, with 72.7 per cent of the samples supporting an excess ($\approx0.6\sigma$).

We also fit the broken-power-law (BPL) and power-law-plus-log-normal (PL+LN) models of Paper II to this profile (Fig.~\ref{fig:tess}). Leave-one-out cross-validation \citep{vehtariPracticalBayesianModel2017} marginally prefers PL+LN, with $|\Delta\mathrm{ELPD}|/\mathrm{SE}_\Delta=1.7$. However, PL+LN does not use its additional flexibility to describe an excess within the ridge: the PL+LN profile reaches its maximum at $P_\mathrm{max}=6.4^{+1.9}_{-1.5}$\,d, just beyond the ridge’s outer edge, so that describing the published ridge would require a further component. The BPL has an inner slope $\alpha_1=2.41^{+0.29}_{-0.26}$ and an outer slope $\alpha_2=-0.15^{+0.48}_{-0.59}$, consistent with zero. A negative median outer slope does not by itself indicate a local overdensity; together with the binned profile, it describes a knee beyond which the population extends with roughly constant $\Gamma$ to longer periods. Neither the contrasts nor the model comparison therefore indicate an overdensity at the ridge per unit $\ln P$.

\section{Discussion}
Our argument focuses on how the occurrence density is represented rather than the data. The region claimed as the ridge nevertheless remains of great interest. It lies at the intersection of Neptunian desert boundaries, where the hot-Jupiters meet the abundant sub-Neptunes, and several physical processes may shape it during planet formation and evolution. 

Distinguishing a genuine excess from a plateau with occurrence rates will require larger samples. In Paper~II, we estimated that separating a log-normal ridge from a broken power law at the $3\sigma$ level would need about four times the current number of TESS candidates in the 6--8\,$R_\oplus$ strip, which the extended TESS mission and PLATO can provide. We recommend that such analyses state the variable in which an overdensity is claimed and test it against an appropriate reference, as discussed in Section~\ref{sec:sim}.

In addition, the analysis of CG26 provides an independent comparison between our TESS-SPOC occurrence rates and earlier Kepler studies. Once the factor $P$ is taken into account, the TESS contrast agrees with all five Kepler estimates (Table~\ref{tab:contrasts}). This reinforces the bin-by-bin agreement between the two surveys found in Paper II and supports the consistency of close-in planet demographics across surveys.

More generally, whether a feature appears as an overdensity, a plateau or a deficit can depend on whether occurrence is expressed per unit $P$ or per unit $\ln P$. We therefore encourage future demographic studies to state this choice and to test any claimed feature against a smooth, physically motivated baseline, such as a power law, ideally calibrated with feature-free simulations (Section~\ref{sec:sim}).

\section{Conclusions}
We have shown that the twofold occurrence excess of the Neptunian ridge reported by CG26 arises from expressing the occurrence density per unit period rather than per unit logarithmic period. Our main conclusions are summarised as follows.

\begin{itemize}
    \item Because the savanna interval adopted by CG26 is twice as wide in period as the ridge, a log-uniform population produces a ridge--savanna contrast of 2 in $\mathrm{d}N/\mathrm{d}P$, consistent with the measured $2.14^{+0.64}_{-0.49}$.
    \item Per unit $\ln P$, the TESS and Kepler contrasts compiled by CG26 fall from 1.6--2.2 to 0.8--1.2, all consistent with unity and hence no excess.
    \item A simulated population that is log-uniform outside the CG24 desert, with no intrinsic excess, reproduces the published ridge in $\mathrm{d}N/\mathrm{d}P$.
    \item Re-measured with TESS-SPOC for 5.5--8.5\,$R_\oplus$, the ridge-to-savanna ratio per unit $\ln P$ is $0.91^{+0.21}_{-0.16}$, and even the most populated ridge bin exceeds the savanna with a one-sided Gaussian equivalent of only $0.60\sigma$. The profile rises steeply at the desert edge and then flattens, as described by a broken power law with an outer slope consistent with zero.
\end{itemize}

\section*{Acknowledgements}
This paper includes data collected by the \textit{TESS} mission. Funding for the \textit{TESS} mission is provided by the NASA Explorer Program. Resources supporting this work were provided by the NASA High-End Computing (HEC) Program through the NASA Advanced Supercomputing (NAS) Division at Ames Research Center for the production of the SPOC data products. This research was funded by the UKRI (Grants ST/X001121/1, EP/X027562/1). AI tools were used to double check for any prior literature on demographic analyses conducted over a linear period and to refine and soften some of the language.





\bibliographystyle{mnras}
\bibliography{ridge} 








\bsp	
\label{lastpage}
\end{document}